\documentclass[a4paper]{spie}  

\usepackage{graphicx,epsfig,url}

\title{Lung infection and normal region segmentation from \\CT volumes of COVID-19 cases}

\author[a,b]{Masahiro ODA}
\author[b]{Yuichiro HAYASHI}
\author[c,d]{Yoshito OTAKE}
\author[e]{Masahiro HASHIMOTO}
\author[f]{Toshiaki AKASHI}
\author[a,b,d]{Kensaku MORI}
\affil[a]{Information Strategy Office, Information and Communications, Nagoya University, Furo-cho, Chikusa-ku, Nagoya, Aichi, 464-8601, Japan}
\affil[b]{Graduate School of Informatics, Nagoya University, Nagoya, Japan}
\affil[c]{Graduate School of Science and Technology, Nara Institute of Science and Technology, Nara, Japan}
\affil[d]{Research Center for Medical Bigdata, National Institute of Informatics, Tokyo, Japan}
\affil[e]{Department of Radiology, Keio University School of Medicine, Tokyo, Japan}
\affil[f]{Department of Radiology, Juntendo University, Tokyo, Japan}

\authorinfo{Further author information: 
(Send correspondence to M. Oda)\\
M. Oda: E-mail: moda@mori.m.is.nagoya-u.ac.jp, Telephone: +81 (0)52 789 5688\\
K. Mori: E-mail: kensaku@is.nagoya-u.ac.jp, Telephone: +81 (0)52 789 5689}

\begin{document} 
\maketitle

\begin{abstract}
This paper proposes an automated segmentation method of infection and normal regions in the lung from CT volumes of COVID-19 patients.
From December 2019, novel coronavirus disease 2019 (COVID-19) spreads over the world and giving significant impacts to our economic activities and daily lives.
To diagnose the large number of infected patients, diagnosis assistance by computers is needed.
Chest CT is effective for diagnosis of viral pneumonia including COVID-19.
A quantitative analysis method of condition of the lung from CT volumes by computers is required for diagnosis assistance of COVID-19.
This paper proposes an automated segmentation method of infection and normal regions in the lung from CT volumes using a COVID-19 segmentation fully convolutional network (FCN).
In diagnosis of lung diseases including COVID-19, analysis of conditions of normal and infection regions in the lung is important.
Our method recognizes and segments lung normal and infection regions in CT volumes.
To segment infection regions that have various shapes and sizes, we introduced dense pooling connections and dilated convolutions in our FCN.
We applied the proposed method to CT volumes of COVID-19 cases.
From mild to severe cases of COVID-19, the proposed method correctly segmented normal and infection regions in the lung.
Dice scores of normal and infection regions were 0.911 and 0.753, respectively.
\end{abstract}


\section{INTRODUCTION}
\label{sec:intro}  

Novel coronavirus disease 2019 (COVID-19) was recognized in December 2019.
It spreads over the world causing the large number of infected patients.
The total numbers of cases and deaths related to COVID-19 are more than 103 million and 2 million in the world by February 1, 2021 \cite{worldmeters}.
Providing appropriate treatments to a patient and prevention of infection based on diagnosis result of the patient are important.
However, because of the rapid increase of COVID-19 patients, medical institutions are suffering from a manpower shortage.
Development of a computer aided diagnosis (CAD) system for COVID-19 is pressing demanded to reduce load on medical staffs.
Reverse transcriptase polymerase chain reaction testing (RT-PCR) is used to diagnose COVID-19.
However, the sensitivity of RT-PCR is not high, ranging from 42\% to 71\% \cite{Simpson20}.
In contrast, the sensitivity of chest CT image-based COVID-19 diagnosis is reported as 97\% \cite{Ai20}.
Chest CT is effective for diagnosis of viral pneumonia including COVID-19.
CT image-base CAD systems are important in COVID-19 diagnosis.
To develop such CAD systems, development of a quantitative analysis method of condition of the lung is required.
Ground-glass opacities (GGOs) and consolidations are commonly found in the lung regions of viral pneumonia cases.
We call regions including GGO and consolidation as {\it infection region}.
The remaining region in the lung is defined as a {\it normal region}.
Quantitative analysis of such regions is useful in lung condition analysis.

Previously, some segmentation methods of infection and normal regions from CT volumes of COVID-19 cases were proposed \cite{infnet,Yan20}.
Inf-Net \cite{infnet} segments infection regions from CT volumes.
The method uses reverse attention and edge attention in their FCN-based segmentation model.
It obtained good segmentation accuracy even from small dataset that contains 45 cases of CT volumes.
Yan et al. \cite{Yan20} proposed a segmentation method of infection and normal regions.
They used progressive atrous spatial pyramid pooling in their FCN model.
They trained the model using large dataset that contains 731 cases of CT volumes.

We propose an automated segmentation method of infection and normal regions in the lung from CT volumes using a fully convolutional network (FCN) for COVID-19 segmentation.
In the segmentation process, both local and global-level features are important to segment infection regions that have various sizes and shapes.
We use an COVID-19 segmentation FCN that has dense pooling connections and dilated convolutions to segment infection regions accurately.
The dense pooling connections enables utilization of multi-scale spatial features that are extracted in the encoder of FCN in the bottleneck layer.
The dilated convolutions also contribute to extract multi-scale spatial features.
These components in the FCN improve segmentation performances of target structures that have large variations in their sizes and shapes.
Also, these components contribute to achieve high generalization ability on the FCN even from small dataset.
It was confirmed in our evaluation using a small dataset.

The contributions of this paper can be summarized as: (1) proposal of the COVID-19 segmentation FCN that has high performance in segmentation of infection regions which have large variations in their sizes and shapes and (2) proposal of the FCN that can achieve high generalization ability from small dataset.

\section{METHOD}
\label{sec:method}

\subsection{Overview}

One application of the proposed method is CAD system for COVID-19 case.
CT volumes of various slice thickness (1.0 to 5.0 mm) are used in diagnosis.
The proposed method is designed to accept various slice thickness by employing a multiple axial slice processing approach.

A CT volume containing the lung region is processed by the proposed method.
The COVID-19 segmentation FCN performs segmentation from multiple axial slices.
Sequence of axial slices obtained from the CT volume is given to the FCN.
The FCN segments infection and normal regions in the lung.
The slice-wise segmentation results are reconstructed as an output volume.

\subsection{Image preparation}

Input of the FCN is a sequence of axial slices obtained from the input CT volume.
All axial slices are scaled to the image size of 384$\times$384 pixels.
CT values in the range from -2050 to 950 H.U. are normalized to the range from -1.0 to 1.0.
Three axial slices of consecutive slice numbers are combined in a three color channel image.
The three color channel image is used as the input of the FCN.

Ground truth images of the region contain three regions including normal region, infection region, and air region outside the body.

\subsection{COVID-19 segmentation FCN}

U-Net \cite{unet} is commonly used in image segmentation.
While the U-Net is effective for segmentation of convex objects such as the lung and liver, small or thin objects are easily missed.
This is because features of small structures are lost in multiple max-pooling operations in the encoding path in the U-Net.
Infection regions have various shapes including small, thin, and complicated shapes.
Infection regions are difficult to segment by using U-Net.
Recent work trying to improve segmentation accuracy of such regions.
Dense pooling connection \cite{playout18} was proposed to utilize spatial information of multiple scales in layers in the encoding path and the bottleneck layer to perform segmentation.
The bottleneck layer in the U-Net performs feature conversions from input to output images.
However, spatial information in the input image is lost in max pooling operations in the encoding path.
The dense pooling connection provides spatial information to the bottleneck layer.
Furthermore, mixed poolings \cite{playout18} are used in the dense pooling connections instead of max poolings to reduce information loss by max pooling operations.
The mixed pooling is implemented as a combination of max and average poolings.
Dilated convolution \cite{Yu16} was proposed to utilize sparsely-distributed features in convolution operations.
It performs a sparse convolution operation from feature maps.
By combining results of dilated convolution operations of multiple dilation rates, we obtain a convolution result of multiple-scales in a feature map.
We utilize these techniques in our segmentation.

\begin{figure}[tb]
\begin{center}
\includegraphics[width=0.98\textwidth]{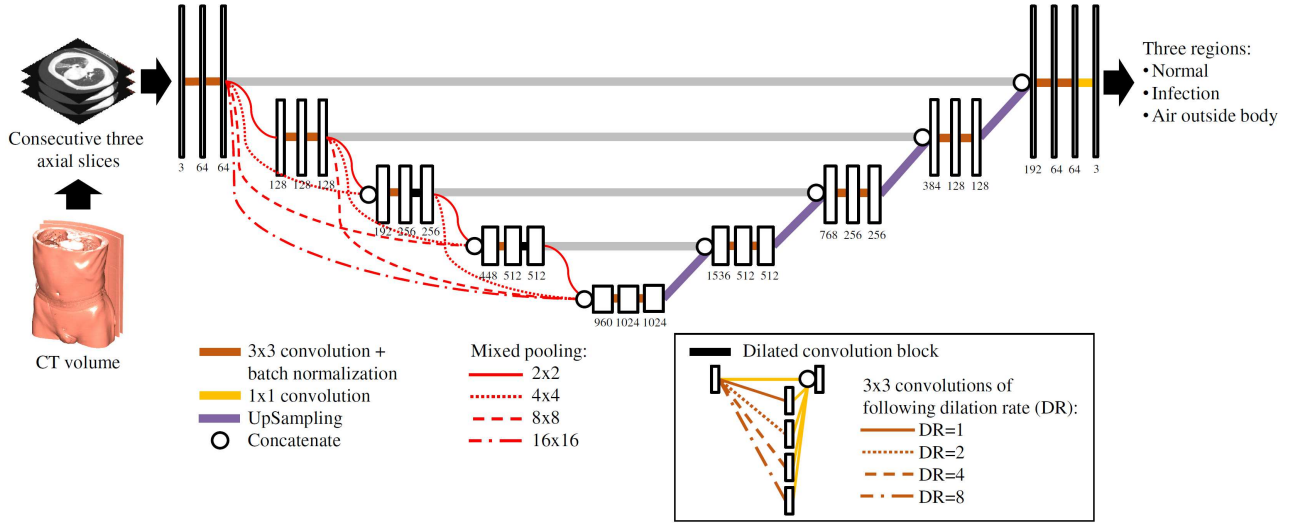}
\end{center}
\caption{COVID-19 segmentation FCN. White boxes are feature maps or images. Numbers below boxes are numbers of kernel or color channel. Dense pooling connections are represented as red connections, which are implemented as combination of mixed poolings. Two dilated convolution blocks are inserted to encoding path. Structure of dilated convolution block is shown in box.}
\label{fig:network}
\end{figure}

The structure of our COVID-19 segmentation FCN is shown in Fig. \ref{fig:network}.
We employ the dense pooling connections in the encoding path of the FCN.
Also, we introduce two dilated convolution blocks in the encoding path.
The dilated convolution block consists of parallelly-connected dilated convolutions of multiple dilation rates.
The results of dilated convolutions are combined using a concatenate operation.

\subsection{Segmentation}

The FCN is trained using CT volumes for training and corresponding ground truth volumes.
In the inference step, images obtained from CT volumes for inference are given to the trained FCN.
Segmentation results of normal and infection regions are reconstructed as the output volume.

\section{EXPERIMENTS AND RESULTS}
\label{sec:experiments}

We applied the proposed method to 20 CT volumes of COVID-19 patients.
The CT volumes were taken in multiple medical institutions in Japan.
Specifications of the CT volumes are: image size was 512$\times$512 pixels, slice number was 56 to 722, pixel spacing was 0.63 to 0.78 mm, and slice thickness was 1.00 to 5.00 mm.
Ground truth volumes were checked by a radiologist.
In the training of the FCN, the number of minibatch was 10 and the training epoch number was 100.
The generalized dice loss \cite{sudre17} was used as a loss function.

We performed a 10-fold cross validation in our evaluation.
The evaluation criterion is the dice score of normal and infection regions.

\begin{figure}[tb]
\begin{center}
\begin{tabular}{c}
\includegraphics[width=0.5\textwidth]{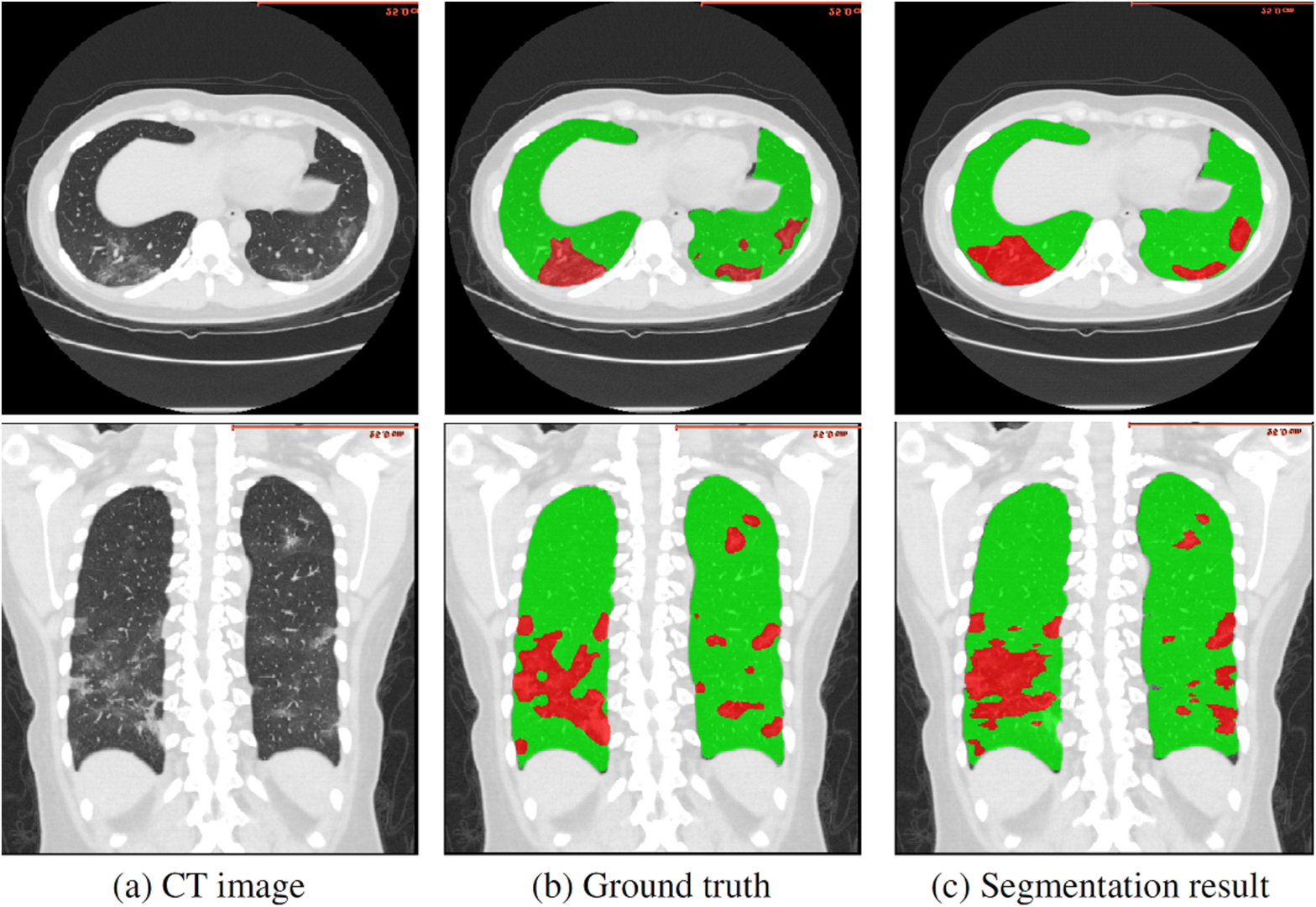}\\
(a)\\
\\
\includegraphics[width=0.5\textwidth]{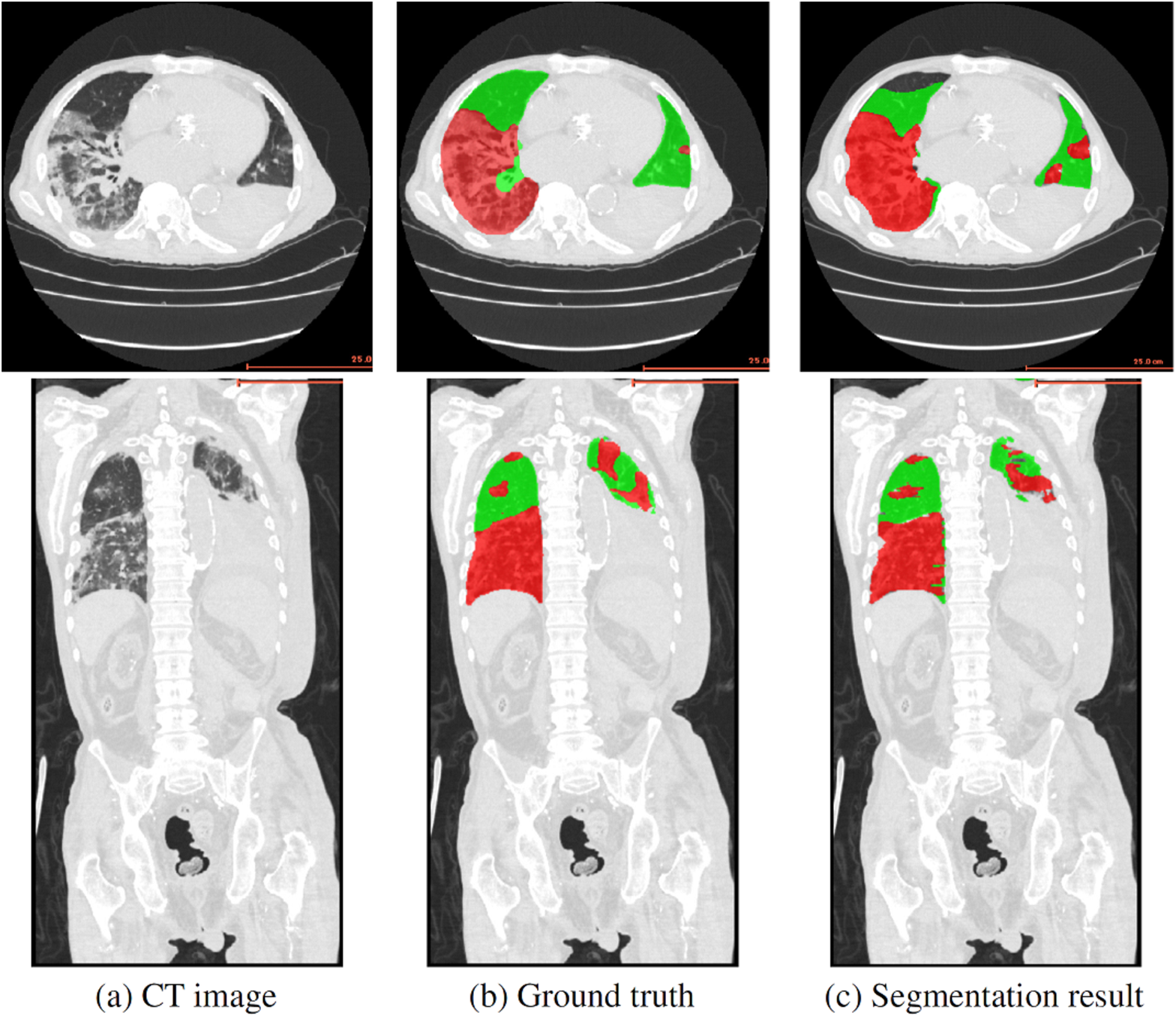}\\
(b)\\
\\
\includegraphics[width=0.5\textwidth]{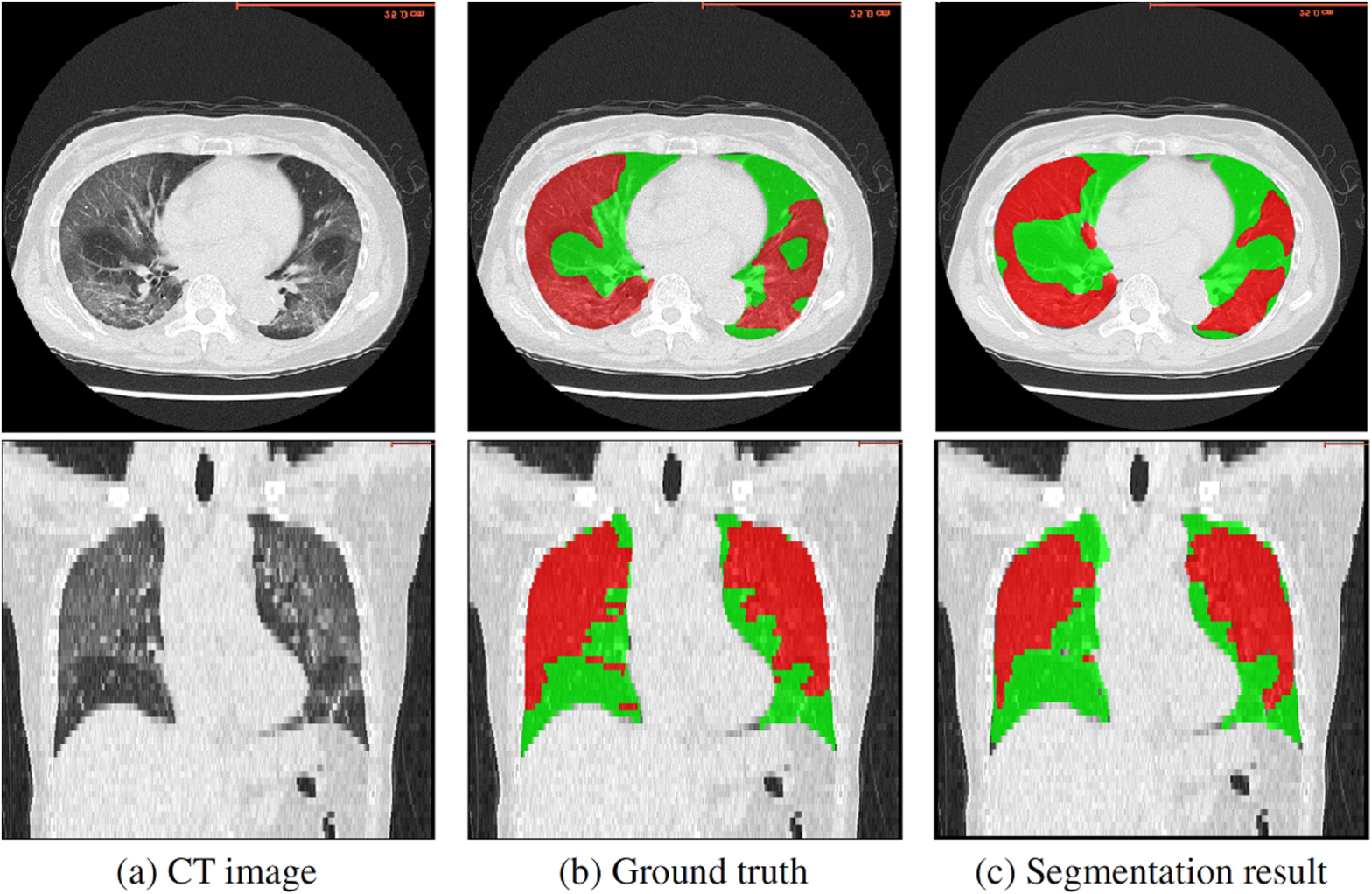}\\
(c)
\end{tabular}
\end{center}
\caption{Results from (a) chest CT of 1.25 mm slice thickness, (b) CT volume of 1.00 mm slice thickness including chest and abdominal regions, and (c) chest CT of 5.00 mm slice thickness. Green and red regions are normal and infection regions. Top and bottom columns show axial and coronal slices.}
\label{fig:result_all}
\end{figure}

Segmentation results are shown in Fig. \ref{fig:result_all}.
The proposed method achieved 0.911 and 0.753 dice scores of normal and infection regions, respectively.
As a comparison, we performed segmentation using U-Net \cite{unet} and obtained 0.898 and 0.741 dice scores of normal and infection regions.
The dice scores were compared with the previously proposed methods \cite{infnet,Yan20} and summarized in Table \ref{tab:comparison}.
Among these methods, the proposed method achieved the highest dice score of infection region.

\begin{table}[tb]
\begin{center}
\caption{Dice scores of proposed and previous segmentation methods for COVID-19 cases. The highest score among each region is indicated by bold text.}
\label{tab:comparison}
\begin{tabular}{|c|c|c|} 
\hline
 & \multicolumn{2}{|c|}{Dice score} \\ \cline{2-3}
Method & Normal region & Infection region \\ \hline
Inf-Net\cite{infnet} & - & 0.739 \\ \hline
Yan et al.\cite{Yan20} & \bf{0.987} & 0.726 \\ \hline
U-Net\cite{unet} & 0.898 & 0.741 \\ \hline
Proposed method & 0.911 & \bf{0.753} \\ \hline
\end{tabular}
\end{center}
\end{table}

\section{DISCUSSION}

The proposed method has a high performance in segmentation of infected region from CT volumes of COVID-19 patients.
It is shown in the Table \ref{tab:comparison}.
The result proofs the proposed method has high generalization ability because the method obtained the highest dice score even from training on the small dataset.
The method proposed by Yan et al. \cite{Yan20} achieved the highest dice score of normal region.
Because they used a large dataset that contains 731 CT volumes, their FCN got good segmentation performance.
The proposed method is expected to achieve higher dice scores than results shown in the Table \ref{tab:comparison} if it is trained on a large dataset.
We will continue working to obtain more data to train the proposed method.

CT volumes contain not only chest but also abdominal regions as shown in Fig. \ref{fig:result_all} (b).
The proposed method successfully segmented lung regions regardless of scan range of CT volumes.
It indicates our 2D axial slice-based process has enough performance to segment 3D lung regions.

The slice thickness of CT volumes varies from 1.00 to 5.00 mm.
Figures \ref{fig:result_all} (b) and (c) show segmentation results from CT volumes of 1.00 and 5.00 mm slice thicknesses.
Even though there is a difference of resolution along the body axis, the proposed method segmented both normal and infection regions.
This is a benefit of employing 2D slice-based process.
3D image-based process is difficult to perform segmentation from CT volumes having inhomogeneous resolutions.

Variation of the shape and intensity of infection regions makes segmentation difficult.
Because we included GGO and consolidation in the infection region, intensities, distribution pattern has large variation.
Therefore, the dice score of infection regions were lower than normal region.
We will improve the segmentation accuracy of infection regions by adding more training data and introducing data augmentation techniques.

\section{CONCLUSIONS}

We proposed an automated segmentation method of infection and normal regions in the lung from CT volumes.
The method utilizes a COVID-19 segmentation FCN that has the dense pooling connections and pararelly-connected dilated convolutions to segment infection regions that have various shapes and intensities.
In the evaluation utilizing 20 CT volumes of COVID-19 patients, we achieved segmentation accuracies of 0.911 and 0.753 in dice scores from normal and infection regions in the lung.

\acknowledgments 
 
Parts of this research were supported by the AMED Grant Numbers 18lk1010028s0401,
JP19lk1010036, and JP20lk1010036, the MEXT/JSPS KAKENHI Grant Numbers 26108006, 17H00867, and 17K20099, the JSPS Bilateral International Collaboration Grants. We used the Japan Medical Image Database (J-MID) created by the Japan Radiological Society with support by the AMED Grant Number JP20lk1010025.

\bibliography{21spie_paper_cite} 
\bibliographystyle{spiebib} 

\end{document}